# Optimization performance of quantum Otto heat engines and refrigerators with squeezed thermal reservoirs

Yanchao Zhang[*]

*School of Science, Guangxi University of Science and Technology , Liuzhou 545006, People's Republic of China*

We consider a quantum Otto cycle operating between two squeezed thermal reservoirs. The influences of the squeezing degree on the optimization performance of quantum Otto heat engines and refrigerators are investigated. We demonstrate that under symmetric condition, the efficiency at maximum power (EMP) of heat engines and the coefficient of performance (COP) at maximum $\chi$ criterion of refrigerators are equal to Curzon-Ahlborn (CA) efficiency and CA COP, respectively. We also found that under asymmetric condition, the EMP of heat engines can be improved when the squeezing degree of hot thermal reservoir is greater than that of the cold thermal reservoir, while be reduced or even inhibited in the opposite condition. However, the COP at maximum $\chi$ criterion of refrigerators can be enhanced when the squeezing degree of cold thermal reservoir is greater than that of the hot thermal reservoir, otherwise will be suppressed.

---

[*] Email: zhangyanchao@gxust.edu.cn

## I. INTRODUCTION

Quantum effects have a significant impact on the performance of quantum thermal machines at the micro-nano scale[1,2]. In particular, theoretical studies have indicated that the efficiency of energy conversion can be increased beyond the standard thermodynamic bounds by employing quantum coherence[3,4], quantum correlation[5,6], quantum phase transition[7], quantum measurement-induced[8], and quantum boundary effect[9]. In recent years, squeezed thermal reservoir as a quantum resource has been widely used in the research of quantum thermodynamics[10-13]. A theoretical work by Roßnagel et al. consider a quantum Otto heat engine coupled to a high temperature squeezed thermal reservoir, while the low temperature thermal reservoir is still purely thermal. They found that the efficiency at maximum power (EMP) can be dramatically enhanced with the squeezing parameter, even surpass the standard Carnot limit[14]. In the following work, Long et al. further studied the performance of a quantum Otto refrigerator coupled to a low temperature squeezed thermal reservoir and high temperature purely thermal reservoir and shown that squeezing can enhance the coefficient of performance (COP) [15]. Recent experiment shows that the efficiency of a nanomechanical engine consisting of a vibrating nanobeam coupled to squeezed thermal noise is not bounded by the standard Carnot limit[16]. Although these studies show that squeezing can improve the performance of the quantum thermal machines, One major open question is whether quantum thermal machines can be improved by quantum squeezing when working between two squeezed thermal reservoirs.

In this paper, we will consider a most general case, which is a quantum Otto cycle coupling to two squeezed thermal reservoirs. We will focus on the influences of squeezing on the optimization performance of quantum Otto heat engines and refrigerators under the $\chi$ figure of merit. Here, the figure of merit $\chi$, defined as $\chi = zQ_{in}/t_{cycle}$, is first proposed by de Tomás et al. as a unified optimization criterion for heat engines and refrigerators[17]. Here $z$ is the converter efficiency, $Q_{in}$ is the heat absorbed by the working system and $t_{cycle}$ is the time duration of cycle. For heat engines $\chi$ becomes power output[18] while for refrigerators it becomes target function $\varepsilon Q_c$ [19]. We found that the optimal efficiency (or COP for refrigerators) is equal to the classical CA value in the symmetric case, and it can be improved only in some specific asymmetric cases, while be reduced or even inhibited in the opposite asymmetric cases.

This paper is organized as follows. In Sec. II, we briefly describe the model and basic physical theory of a quantum Otto cycle with squeezed thermal reservoirs. In Sec. III and IV, we investigate the influences of the squeezing degree on the EMP of quantum Otto heat engines and COP at maximum $\chi$ criterion of refrigerators, respectively. We summarize our results in Sec. V.

## II. QUANTUM OTTO CYCLE WITH SQUEEZED THERMAL RESERVOIRS

We consider a quantum Otto cycle whose working medium is a single harmonic oscillator with time-dependent frequency $\omega$ [20-24]. The statistical and thermodynamical properties of a harmonic oscillator with time-dependent frequency $\omega$ coupled to a squeezed thermal reservoir is described in a position-momentum frame $(x_0, p_0)$. In this frame, the position quadrature $x_0$ and the momentum quadrature $p_0$ corresponds to the anti-squeezed quadrature and the squeezed quadrature, respectively. The probability of the harmonic oscillator at position $x_0$ obeys the Gaussian distribution with temperature $T_1$, whereas the probability of the harmonic oscillator at momentum $p_0$ obeys the Gaussian distribution with temperature $T_2$ [16]

$$\rho(x_0) = \sqrt{\frac{\hbar\omega}{2\pi k_B T_1}} \exp\left(-\frac{\hbar\omega x_0^2}{2k_B T_1}\right), \tag{1}$$

$$\rho(p_0) = \sqrt{\frac{\hbar\omega}{2\pi k_B T_2}} \exp\left(-\frac{\hbar\omega p_0^2}{2k_B T_2}\right). \tag{2}$$

Where $T_1$ and $T_2$ are two temperatures that describe the level of fluctuations in the anti-squeezed and squeezed quadratures. These temperatures can be used to define the effective temperature $T$ and squeezing parameter $r$ to describe the squeezed thermal reservoirs via $T_{1,2} = T\exp(\pm 2r)$. As a result, the internal energy of the harmonic oscillator is given by [16]

$$\begin{aligned} U &= \int_{-\infty}^{+\infty}\int_{-\infty}^{+\infty} \rho(x_0)\rho(p_0)\left(\frac{\hbar\omega}{2}\right)(x_0^2 + p_0^2)dx_0 dp_0 \\ &= k_B T \cosh(2r) \end{aligned} \tag{3}$$

The quantum Otto cycle consists of two isentropic processes and two isochoric processes, as shown in Fig. 1.

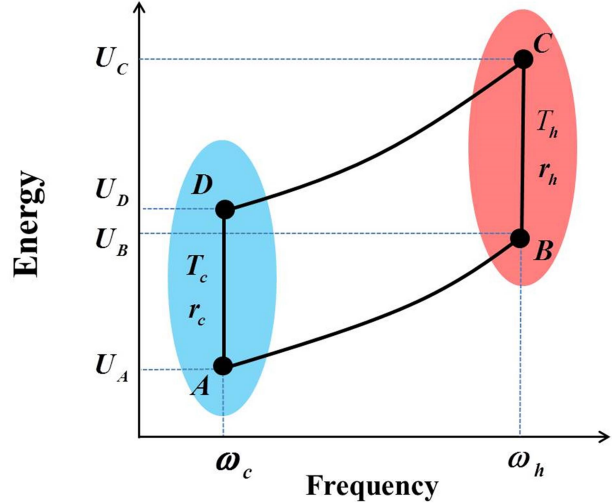

FIG. 1. Energy-frequency diagram of the quantum Otto cycle with squeezed thermal reservoirs.

During the isochoric processes, the system is alternatingly coupled to a hot squeezed thermal reservoirs with effective temperature $T_h$ and squeezed parameter $r_h$ and a cold squeezed reservoir with temperature $T_c(<T_h)$ and squeezed parameter $r_c$. In these processes, the frequency is constant, and thus, only heat is exchanged with the squeezed thermal reservoirs. During the isentropic processes, the frequency of the oscillator is modulated between $\omega_c$ and $\omega_h$ ($\omega_h > \omega_c$) while the system is isolated.

Thus, work is done by the system in these processes. The isentropic processes are implemented by simultaneously varying frequency and temperature such that the ratios $T/\omega$ remain constant. In this case, the entropy of the system is constant. Thus, the four squeezed thermal states of quantum Otto cycle are $A(T_c, r_c)$, $B(T_c\omega_h/\omega_c, r_c)$, $C(T_h, r_h)$ and $D(T_h\omega_c/\omega_h, r_h)$, respectively. The internal energies of these squeezed thermal states are given by

$$U_A = k_B T_c \cosh(2r_c), \tag{4}$$

$$U_B = k_B T_c \frac{\omega_h}{\omega_c} \cosh(2r_c), \tag{5}$$

$$U_C = k_B T_h \cosh(2r_h), \tag{6}$$

$$U_D = k_B T_h \frac{\omega_c}{\omega_h} \cosh(2r_h). \tag{7}$$

## III. QUANTUM OTTO HEAT ENGINES AT MAXIMUM POWER

For a quantum Otto heat engine, the thermodynamic cycle follows counterclockwise, i.e. $A \to B \to C \to D \to A$, as shown in Fig. 1. It can be described by the following four processes: (1) Isentropic compression process: In this process, the system is isolated and the system states from $A(T_c, r_c)$ to $B(T_c\omega_h/\omega_c, r_c)$ the frequency increases from $\omega_c$ to $\omega_h$. This transformation is unitary and the entropy is constant. The mean input work done during this process is given by $W_{in} = U_B - U_A$. (2) Isochoric heat addition process: The system is coupled to a hot squeezed thermal reservoir. We assume the duration of the isochoric process to be much shorter than that of the isentropic process. Thus, the frequency stays constant at fixed $\omega_h$ in this process. As a result, the system relaxes to squeezed thermal state $C(T_h, r_h)$. In this process, the system absorbs heat from the hot squeezed thermal reservoir and the mean heat is given by $Q_h = U_C - U_B$. (3) Isentropic expansion process: the system decouples from the hot squeezed thermal reservoir and the frequency is changed back to its initial value $\omega_c$ during this process. The isolated system evolves unitarily into state $D(T_h\omega_c/\omega_h, r_h)$ at constant von Neumann entropy. In this process, the system output work $W_{out} = U_C - U_D$. (4) Isochoric heat rejection process: the system is then coupled to a cold squeezed thermal reservoir and quickly relaxes to the initial state $A(T_c, r_c)$. In this process, the frequency $\omega_c$ is kept constant and the heat $Q_c = U_D - U_A$ is released to the cold squeezed thermal reservoir.

After a whole cycle, the system recovers its initial state, the total output work is

$$W_{total} = W_{out} - W_{in} = U_C - U_D - (U_B - U_A)$$
$$= \left(1 - \frac{\omega_c}{\omega_h}\right) k_B T_h \cosh(2r_h) - \left(\frac{\omega_h}{\omega_c} - 1\right) k_B T_c \cosh(2r_c), \tag{8}$$

The efficiency defined as the ratio of the total output work per cycle and the heat from the hot squeezed thermal reservoir, i.e.,

$$\eta = \frac{W_{total}}{Q_h} = 1 - \frac{\omega_c}{\omega_h}. \tag{9}$$

One of the main points of this paper is to study the efficiency of quantum Otto heat engine under the condition of maximum power. The power is given by the total output work divided by the cycle time, i.e. $P = W_{total}/t_{cycle}$. We can obtain the maximum power through the maximum work at a given time with respect to the frequency ratio $\omega_c/\omega_h$, i.e. $dW_{total}/d(\omega_c/\omega_h) = 0$. We found that the power is maximum when the frequency ratio satisfy:

$$\left(\frac{\omega_c}{\omega_h}\right)_{mp} = \sqrt{\frac{T_c \cosh(2r_c)}{T_h \cosh(2r_h)}}, \tag{10}$$

As a result, the EMP is given by

$$\eta_{mp} = 1 - \sqrt{\frac{T_c \cosh(2r_c)}{T_h \cosh(2r_h)}}. \tag{11}$$

The EMP depends explicitly on the squeezing degree. It can be found from Eq. (11)

that in the case of $r_c = r_h = 0$ (for pure thermal reservoirs), the EMP recover the CA efficiency, $\eta_{CA} = 1 - \sqrt{T_c/T_h}$ [18]. In the case where the hot thermal reservoir is squeezed ($r_h \neq 0$), while the cold thermal reservoir is purely thermal reservoir ($r_c = 0$), the EMP recover the result in the Ref. [14]. In particular, we find that in the symmetric case $r_c = r_h$, the EMP is independent of the squeezed parameter and is always equal to CA efficiency, which is one of the main conclusions of this paper. Fig. 2 shows the EMP $\eta_{mp}$ as a function of the temperature ratio $T_c/T_h$ for different squeezed parameters. It is found that when $r_c = 0.5, r_h = 1$, the EMP is obviously greater than CA efficiency, even greater than zero at temperature ratio $T_c/T_h = 1$. This indicates that work can be extracted from a single squeezed thermal reservoir, and detailed studies have been discussed in Ref. [16]. However, in the case of $r_c = 1, r_h = 0.5$, the EMP is significantly less than CA efficiency and it decreases rapidly as the temperature ratio increases.

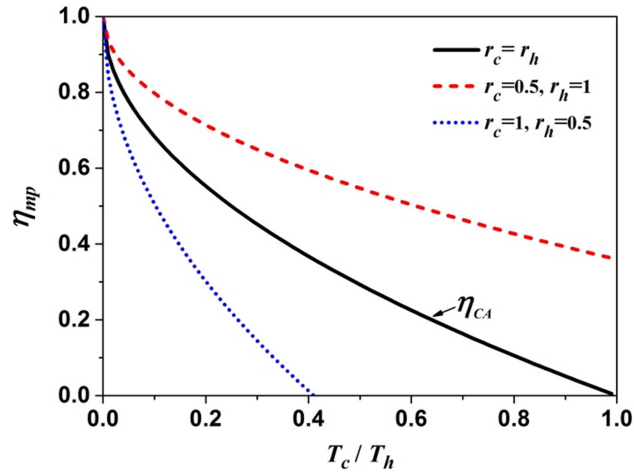

FIG. 2. Efficiency at maximum power $\eta_{mp}$ as a function of the temperature ratio $T_c/T_h$ for different squeezed parameters.

The variation of EMP $\eta_{mp}$ with the squeezed parameters $r_c$ and $r_h$ for fixed temperature ratio $T_c/T_h = 0.25$ is explained in detail, as shown in Fig. 3. The diagonal line (black line), i.e. $r_c = r_h$, corresponds to the CA efficiency $\eta_{CA}$. It is found that $\eta_{mp} > \eta_{CA}$ when $r_h > r_c$, while $\eta_{mp} < \eta_{CA}$ when $r_h < r_c$. This means that the EMP can be enhanced only if the squeezing degree of hot thermal reservoir is greater than that of the cold thermal reservoir, otherwise it will be reduced or even inhibited in the opposite condition. This is the other main conclusion of this paper.

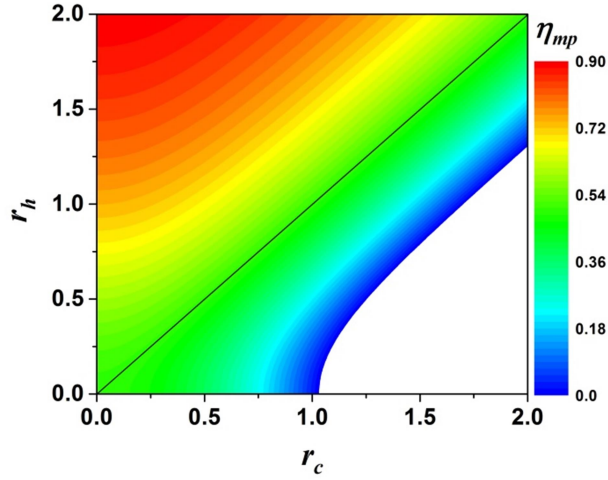

FIG. 3. Efficiency at maximum power $\eta_{mp}$ as a function of the squeezed parameters $r_c$ and $r_h$ for fixed temperature ratio $T_c/T_h = 0.25$.

## IV. QUANTUM OTTO REFRIGERATORS AT MAXIMUM $\chi$ CRITERION

The quantum Otto refrigerator cycle is reverse operation of the quantum Otto heat engine cycle, in other words, it goes through a clockwise cycle process, i.e. $A \to D \to C \to B \to A$. In a refrigeration cycle, the heat exchange processes and the work processes in the heat engine cycle will be completely reversed. Therefore, the heat absorbed from the cold squeezed thermal reservoir is $Q_c = U_D - U_A$, and that released to the hot squeezed thermal reservoir is $Q_h = U_C - U_B$. The total input work in a cycle is given by $Q_h - Q_c$. In order to describe the performance of the refrigerator, the cooling power per unit time and the coefficient of performance are, respectively,

given by

$$Q_c = U_D - U_A$$
$$= \frac{\omega_c}{\omega_h} k_B T_h \cosh(2r_h) - k_B T_c \cosh(2r_c), \quad (12)$$

and

$$\varepsilon = \frac{Q_c}{Q_h - Q_c} = \frac{\omega_c}{\omega_h - \omega_c}. \quad (13)$$

Based on the Eqs. (12) and (13), the figure of merit $\chi = \varepsilon Q_c$, and the maximum $\chi$ is found by setting the derivatives of $\chi$ with respect to frequency ratio $\omega_c/\omega_h$ equal to zero, i.e. $d\chi/d(\omega_c/\omega_h) = 0$, in which case the frequency ratio satisfy the relation

$$\left(\frac{\omega_c}{\omega_h}\right)_{m\chi} = 1 - \sqrt{1 - \frac{T_c}{T_h} \frac{\cosh(2r_c)}{\cosh(2r_h)}}, \quad (14)$$

As a result, the COP at maximum $\chi$ criterion is given by

$$\varepsilon_{m\chi} = \frac{1}{\sqrt{1 - \frac{T_c}{T_h} \frac{\cosh(2r_c)}{\cosh(2r_h)}}} - 1. \quad (15)$$

For pure thermal reservoirs ($r_c = r_h = 0$), the COP at maximum $\chi$ criterion recover the CA COP, $\varepsilon_{CA} = \left(1/\sqrt{1 - T_c/T_h}\right) - 1$ [17,19]. In the case of $r_h = 0$, while $r_c \neq 0$, the result is consistent with that in Ref. [15]. In addition, For symmetric squeezing $r_c = r_h$, the $\varepsilon_{m\chi}$ is not affected by squeezed parameters and is always equal to CA COP. The variation curves of $\varepsilon_{m\chi}$ are plotted as a function of the temperature ratio $T_c/T_h$ for different squeezed parameters, as shown in Fig. 4. It is found that $\varepsilon_{m\chi}$ is obviously less than $\varepsilon_{CA}$ when $r_c = 0.25$, $r_h = 0.5$, and is obviously larger than $\varepsilon_{CA}$ when $r_c = 0.5$, $r_h = 0.25$.

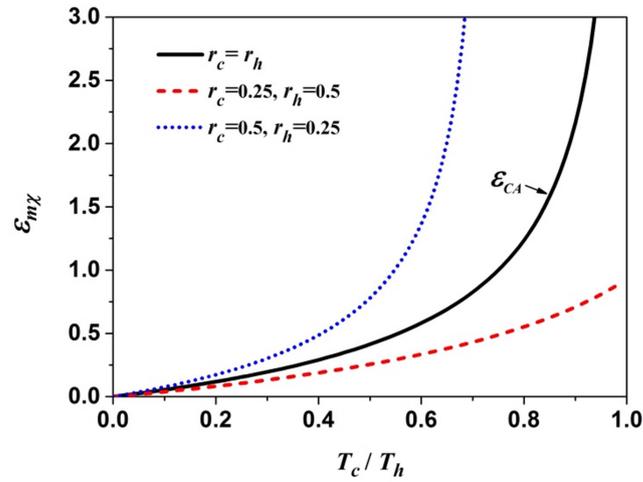

FIG. 4. COP at maximum $\chi$ criterion $\varepsilon_{m\chi}$ as a function of the temperature ratio $T_c/T_h$ for different squeezed parameters.

Fig. 5 shows a more explicit description of the variation of $\varepsilon_{m\chi}$ with the squeezed parameters $r_c$ and $r_h$ for fixed $T_c/T_h = 0.75$. The diagonal line (black line) corresponds to the CA COP, i.e. $\varepsilon_{CA}$. It is shown that the $\varepsilon_{m\chi}$ be enhanced when $r_c > r_h$, while be suppressed when $r_c < r_h$.

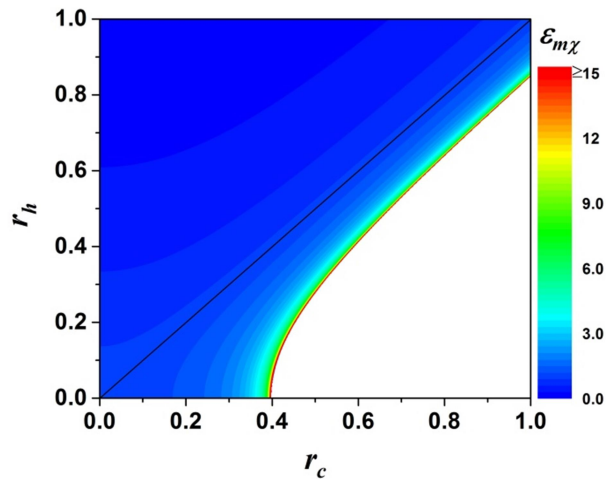

FIG. 5. COP at maximum $\chi$ criterion $\varepsilon_{m\chi}$ as a function of the squeezed parameters $r_c$ and $r_h$ for fixed temperature ratio $T_c/T_h = 0.75$.

## V. CONCLUSIONS

We have studied the optimization performance of quantum Otto heat engines and refrigerators operating between two squeezed thermal reservoirs and clarified the influence of the squeezing degree on the optimization performance. We found that under the symmetric condition, the EMP of heat engines and the COP at maximum $\chi$ criterion of refrigerators are independent of the squeezed parameters and are always equal to CA efficiency and CA COP, respectively. However, under asymmetric condition, the EMP can be enhanced only if $r_h > r_c$, otherwise it will be reduced or even inhibited in the case of $r_h < r_c$, and the COP at maximum $\chi$ criterion can be improved only if $r_c > r_h$, while be suppressed when $r_c < r_h$. The results in this paper show that squeezed thermal reservoir as a quantum resource can improve the optimization performance of the quantum thermal machines only in some specific cases. Whether the same situation exists for other quantum effects (e.g. quantum coherence or quantum correlation) is worthy of further study, which will expand our knowledge in quantum thermodynamics.

**ACKONWLEDGMENTS**

This work is supported by the Guangxi University of Science and Technology Foundation for PhDs (No. 18Z11).